\documentclass[prl,twocolumn,nofootinbib,floatfix,subeqn,showpacs]{revtex4-1}
\usepackage[usenames]{color}
\usepackage{amsmath}
\usepackage{amssymb}
\usepackage{float}
\usepackage{graphicx}
\usepackage{xspace}
\usepackage{hyperref}

\newcommand{\alat}{a_{\rm lat}}

\newcommand{\um}{\,\mu\mbox{m}}

\long\def\symbolfootnote[#1]#2{\begingroup%
\def\thefootnote{\fnsymbol{footnote}}\footnotetext[#1]{#2}\endgroup}

\begin{document}

\title{Coherent light scattering from a two-dimensional Mott insulator}

\author{Christof Weitenberg$^{1}$}
\author{Peter Schau\ss$^{1}$}
\author{Takeshi Fukuhara$^{1}$}
\author{Marc~Cheneau$^{1}$}
\author{Manuel Endres$^{1}$}
\author{Immanuel Bloch$^{1,2}$}
\author{Stefan Kuhr$^{1,3}$}\email{stefan.kuhr@mpq.mpg.de}

\date{18 February 2011}


\affiliation{
   $^1$Max-Planck-Institut f\"ur Quantenoptik, Hans-Kopfermann-Str.~1, 85748 Garching,
   Germany\\
   $^2$Ludwig-Maximilians-Universit\"at, Schellingstr.~4/II, 80799 M\"unchen, Germany\\
   $^3$Department of Physics, Scottish Universities Physics Alliance, University of Strathclyde, Glasgow, UK
}

\begin{abstract}
We experimentally demonstrate coherent light scattering from an atomic Mott insulator
in a two-dimensional lattice. The far-field diffraction pattern of small clouds
of a few hundred atoms was imaged while simultaneously laser cooling the atoms with the probe beams.
We describe the position of the diffraction peaks and the scaling of the peak parameters
by a simple analytic model. In contrast to Bragg scattering, scattering from a single plane
yields diffraction peaks for any incidence angle. We demonstrate the feasibility of detecting
spin correlations via light scattering by artificially creating a one-dimensional
antiferromagnetic order as a density wave and observing the appearance of additional diffraction peaks.
\end{abstract}

\maketitle



Ultracold atoms in optical lattices have become a useful tool to simulate
static phases and the dynamical responses of quantum many-body systems \cite{Bloch:2008c}.
Recent interest has focused on reaching sufficiently low temperatures and entropies
to observe magnetically ordered quantum phases \cite{Lewenstein:2007}. In this context, light scattering has been
proposed as a new tool to detect these quantum correlations. Spin correlations could be mapped onto correlations of scattered light \cite{deVega:2007,*Eckert:2008} or be detected via diffraction peaks from the additional scattering planes for spin-dependent probe light \cite{Corcovilos:2010}.
Light scattering would allow to measure the temperature of fermions in an optical lattice \cite{Ruostekoski:2009}
or the density fluctuations across the superfluid-to-Mott-insulator transition \cite{Mekhov:2007,Lakomy:2009,*Rist:2010}.
Since the amount of scattered light is usually very small, several proposals involve a
cavity for the detection \cite{Mekhov:2007a,*Chen:2007}. Without cavities,
elastic Bragg scattering has been used to demonstrate the long range periodic order of thermal atoms
in an optical lattice despite very low filling factors \cite{Birkl:1995,Weidemueller:1995,*Weidemueller:1998}.
It allowed the measurement of a change of the lattice constant from the backaction of the atoms
\cite{Birkl:1995,Weidemueller:1995,*Weidemueller:1998}, their localization dynamics
\cite{Westbrook:1997,*Raithel:1997} and temperature \cite{Weidemueller:1995,*Weidemueller:1998}.
Bragg scattering was also studied in a far-detuned one-dimensional lattice \cite{Slama:2005a}.

Here we show coherent light scattering from an atomic Mott insulator
(MI) in a two-dimensional square lattice structure. Scattering
from a 2D geometry differs significantly from usual Bragg
scattering, because the momentum transfer needs to be an integer
multiple of a reciprocal lattice vector only within the plane.
Therefore diffraction peaks appear for any angle of the incoming
beam, which is experimentally more convenient.
In our setup, we use five probe beams in a molasses configuration
that simultaneously laser cool the atoms. Each of these molasses beams
yields distinct diffraction peaks in the far-field images.
We quantitatively compared the diffraction patterns with model
calculations and confirmed the coherent nature of the scattering
process. We artificially prepared 1D antiferromagnetic order as a density wave
and observed additional diffraction peaks, thus demonstrating the usability of
light scattering for the detection of global spin correlations.

\begin{figure}[!t]
    \begin{center}
        \includegraphics[width=\columnwidth]{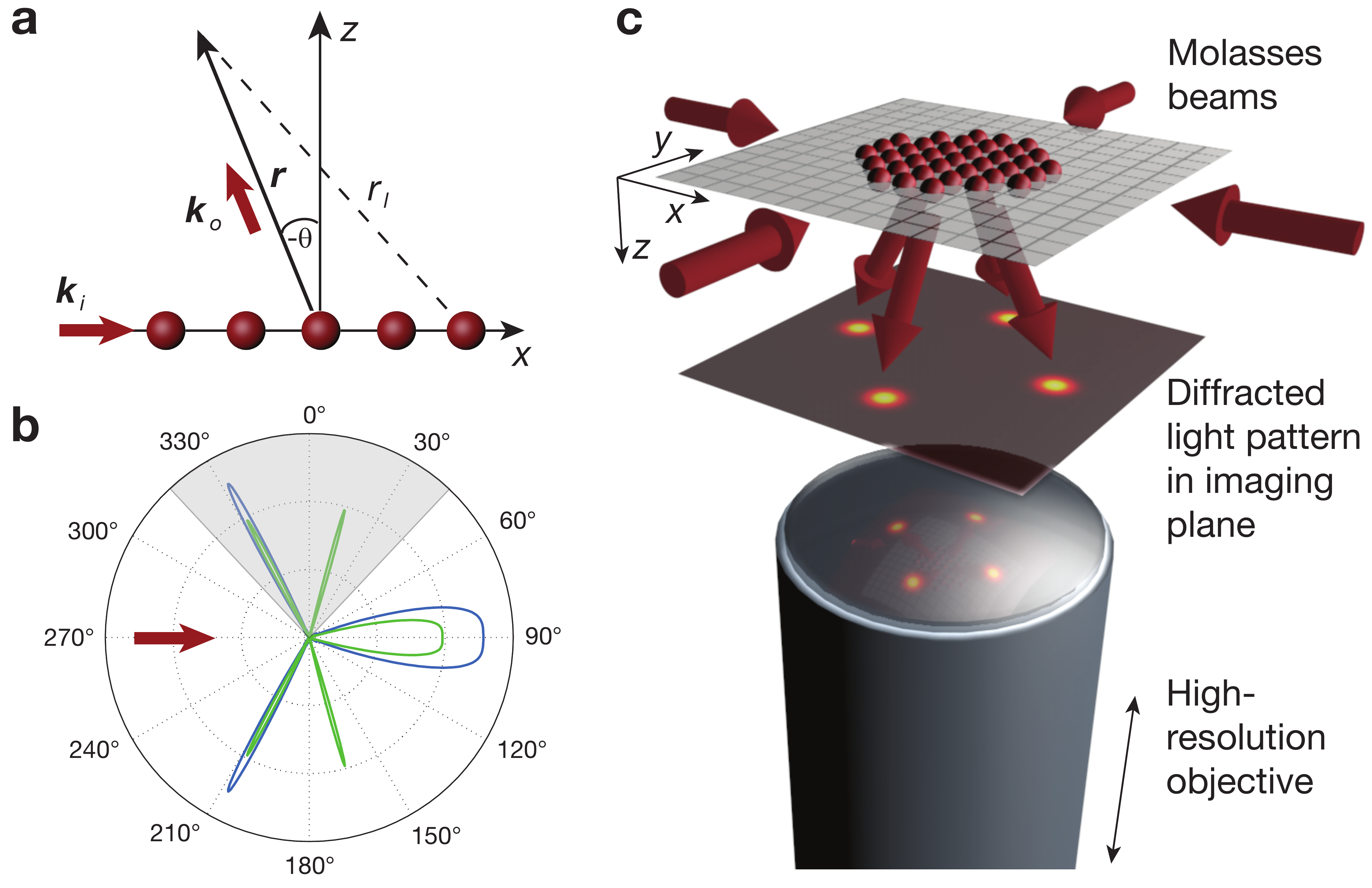}
     \end{center}
     \vspace{-0.5cm}
\caption{Schematics of light scattering. (a) Light diffraction from atoms in a 1D lattice. An
incoming wave with wave vector $\mathbf{k}_i$ is diffracted under an
angle $\theta$. (b) Resulting differential scattering cross section
$\frac{d\sigma}{d\theta}(\theta)$, as given by
Eq.\,(\ref{eq:AngularDependence}) for our experimental parameters.
Shown are the cases for unity filling of the lattice and atom number
$N=16$ (blue curve) and for a $z$-N\'{e}el antiferromagnet along one direction
($N=28$), from which only one spin component is detected (green curve). The grey shaded area
indicates the opening angle of our imaging system.
(c) Experimental setup. Atoms in a 2D optical lattice are
illuminated with four in-plane molasses beams. In situ and far-field
diffraction images are recorded with a high-resolution optical
imaging system. \label{fig:scheme}}\vspace{-0.3cm}
\end{figure}
We begin by introducing a simple analytic 1D model to illustrate the underlying physics.
Atoms on a 1D lattice (lattice spacing $\alat=\lambda_{\rm lat}/2$, $\lambda_{\rm
lat}$ is the lattice wavelength) are driven by an incoming light
field (wavelength $\lambda$) from the $x$ direction
[Fig.\,\ref{fig:scheme}(a)] with wave vector
$\mathbf{k}_{i}=k\,\mathbf{e}_x$, where $k=2\pi/\lambda$ and $\mathbf{e}_x$
is the unit vector along $x$. The scattered light is detected at a
point $\mathbf{r}$, defined by the angle $\theta$ with the $z$-axis,
such that $\mathbf{r}=r\,\mathbf{e}_r=r \sin\theta \,\mathbf{e}_x + r \cos\theta \,\mathbf{e}_z$.
The position of the $l$-th atom is $\mathbf{x}_l=l \,\alat \,\mathbf{e}_x$
and its distance $r_l$ to the detection point is in far-field
approximation $r_l=|\mathbf{r}-\mathbf{x}_l| \approx
r-\mathbf{x}_l \cdot \mathbf{e}_r$. In our model, each atom emits a
spherical wave, which at the detection point can be written as
\begingroup
    \setlength{\abovedisplayskip}{6pt}
    \setlength{\belowdisplayskip}{6pt}
    \begin{equation}
 F_l(r_l)
 =
 f \frac{e^{i k r_l}}{r_l} e^{i \delta_l}
 \approx f \frac{e^{i k r}}{r} e^{-i\mathbf{K}\cdot \mathbf{x}_l}.
    \end{equation}
\endgroup
Here, $f$ denotes the coherently scattered field amplitude, $\delta_l = \mathbf{k}_i \cdot
\mathbf{x}_l$ is the phase imprinted by the incoming light field, and
$\mathbf{K}=\mathbf{k}_{o}-\mathbf{k}_{i}$ with the wave vector
$\mathbf{k}_{o}=k \,\mathbf{e}_r$ in the observed direction. The
differential cross section $\frac{d\sigma}{d\theta}(\mathbf{K})\propto |\sum_l
e^{-i \mathbf{K}\cdot\mathbf{x}_l}|^2$ is obtained by summing over the
field amplitudes from all $N_x$ atoms. As a result, we obtain the
angular dependence of the scattered light field,
\begingroup
    \setlength{\abovedisplayskip}{6pt}
    \setlength{\belowdisplayskip}{6pt}
    \begin{equation}
\frac{d\sigma}{d\theta}(\theta)
\propto \frac{\sin^2 \left[k \,\alat(\sin
\theta-1)N_x/2\right]}{\sin^2\left[k \,\alat(\sin \theta-1)/2\right]}.
    \label{eq:AngularDependence}
    \end{equation}
\endgroup
with distinct maxima when the field
amplitudes of neighboring atoms interfere constructively, i.e.\!
$\mathbf{K}\cdot (\mathbf{x}_l-\mathbf{x}_{l+1})=2\pi\cdot n$, where $n$ is an
integer that denotes the diffraction order. The height of the
diffraction peak is proportional to $N_x^2$ whereas the peak width
scales as $1/N_x$. The angles $\theta_n$, under which the
diffraction maxima can be observed, are given by
\begingroup
    \setlength{\abovedisplayskip}{4pt}
    \setlength{\belowdisplayskip}{4pt}
    \begin{equation}
 \sin{\theta_n}=1+n\frac{\lambda}{\alat}.
    \label{eq:DiffractionCondition}
    \end{equation}
\endgroup
The trivial case $n=0$ gives the forward scattered light ($\theta_0
= 90^\circ$), independent of $\alat$ and $\lambda$. For our
experimental parameters ($\lambda=780\,$nm,  $\alat=532\,$nm),
Eq.\,(\ref{eq:DiffractionCondition}) can be additionally fulfilled only for
$n=-1$, yielding the corresponding minus first diffraction order at
$\theta_{-1}=-27.8^\circ$ and $207.8^\circ$.
These two out of plane scattered waves ensure the momentum conservation in the $z$ direction.
Fig.\,\ref{fig:scheme}(b) (blue curve) shows a polar plot of
$\frac{d\sigma}{d\theta}(\theta)$, displaying the forward scattered
light and the two peaks, one of which is captured by our imaging
system (gray shaded region). If only  every second lattice site is
occupied (e.g.~after removing one spin component in an
antiferromagnetically ordered sample), the periodicity of the system is doubled.
In this case, there are two possible diffraction orders (in the upper half plane)
at $\theta_{-1}^{\rm AFM}=15.5^\circ$ and $\theta_{-2}^{\rm
AFM}=\theta_{-1}=-27.8^\circ$ [green curve in Fig.\,\ref{fig:scheme}(b)].

In our experiment, we prepared 2D MIs of $^{87}$Rb atoms
in an optical lattice. The atoms were confined in a single antinode
of a vertical standing wave. Two pairs of horizontal laser beams
provided the square
lattice structure (see Ref.~\cite{Sherson:2010}). In order to detect
the atoms in the optical lattice, the lattice depth was increased to
$\sim 300\,\mu$K, thereby freezing the atom distribution. The atoms where then
illuminated with an optical molasses, which consisted of two pairs
of retroreflected laser beams oriented along the horizontal lattice axes
[see Fig.\,\ref{fig:scheme}(c)].
A fifth molasses beam entered from the reverse direction of the imaging
system (not shown in Fig.\,\ref{fig:scheme}(c)). The molasses was red detuned with respect to the free space
resonance by 45\,MHz and the total scattering rate was 60\,kHz. We
detected the fluorescence photons with a high resolution microscope
objective with half opening angle $\alpha=43^\circ$. The objective
can be moved in the $z$ direction by a distance $\Delta z<100\,\mu$m between the atom position and the focal plane
within 50\,ms using a piezo scanning device.

%
%
\begin{figure}[b]
    \begin{center}
        \includegraphics[width=\columnwidth]{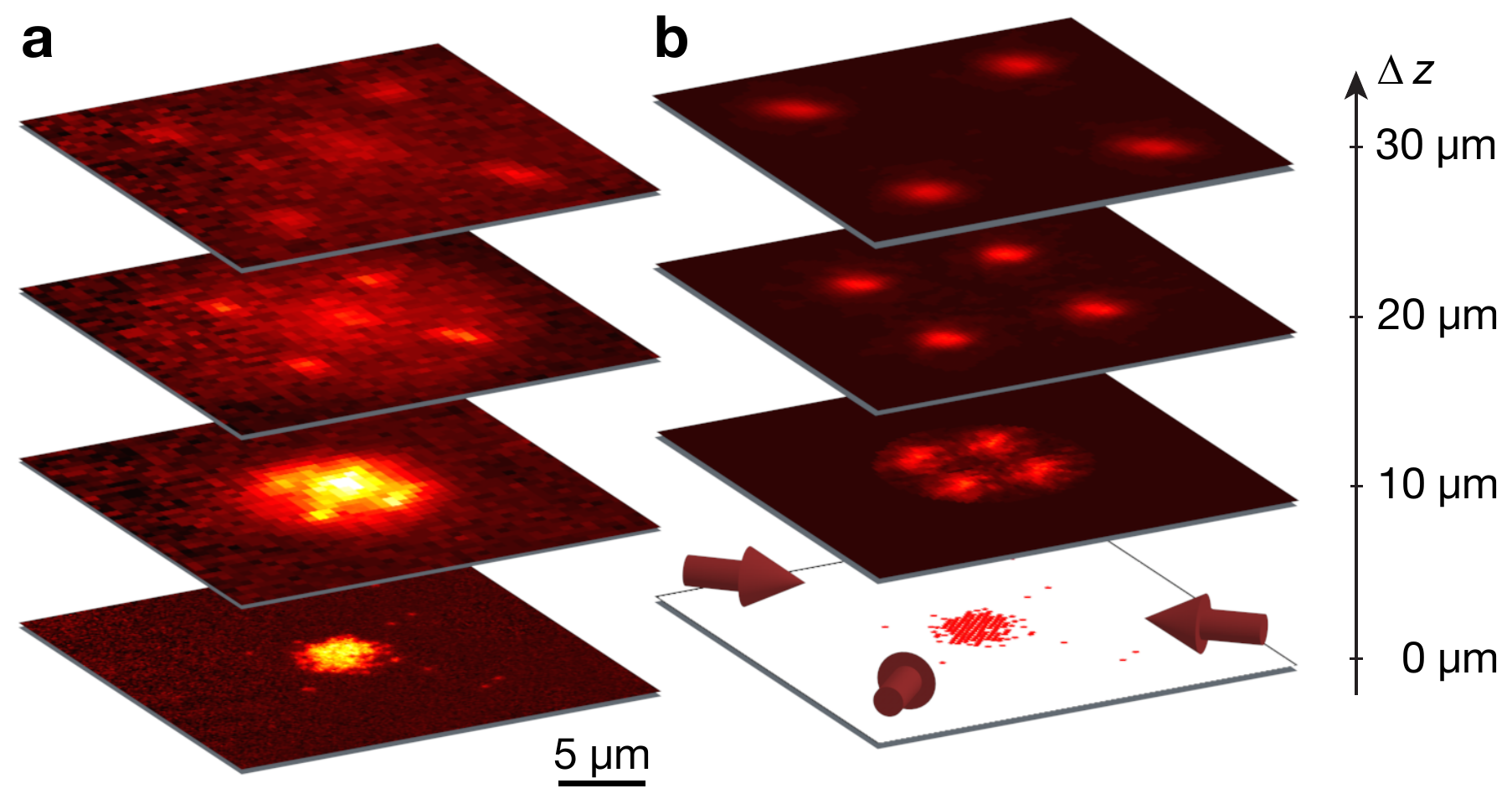}
    \end{center}
    \vspace{-0.5cm}
\caption{Light scattering from a 2D MI. (a) Experimental
images from the same atomic sample for four different distances
$\Delta z$ of the focal plane with respect to the atom position.
The lowest image shows the in situ atom number
distribution, whereas the upper image shows the far-field
diffraction pattern. (b) Simulated diffraction patterns obtained
from a 2D numerical model (see text for details), using the
reconstructed atom distribution (lower image) from (a). Red arrows
indicate the directions of the optical molasses beams. \label{fig:StepFocus}}
\end{figure}

Our experimentally obtained diffraction images are shown in
Fig.\,\ref{fig:StepFocus}(a) for four different distances $\Delta z$
and an illumination time of 200\,ms for each image. For $\Delta
z=0$ we observed the in situ atom number distribution, consisting in
this case of 147 atoms in a MI shell with unit occupancy and diameter of
$6\,\mu$m. For larger $\Delta z$, we observe the build up of the far-field
distribution with distinct diffraction peaks.
We compared the experimental data with a numerical calculation of
$\frac{d\sigma}{d\theta}(\mathbf{K})$ using
the actual atom distribution of the image at $\Delta z = 0$
[Fig.\,\ref{fig:StepFocus}(b)]. For this purpose, we coherently
summed over all spherical waves $F_l(r_l)$ emitted by the atoms with
phases $\delta_l$ given by the incident driving fields. Our model
assumes that all four horizontal molasses beams are diffracted
independently.
A spherical wave for the emission pattern is used, because the
different local polarizations in the molasses result
in all possible orientations of the atomic dipole.
The calculated far-field distribution is in good qualitative
agreement with the experimental images. Our simulation
only includes the coherently scattered light, whereas the experimental data
also shows a significant incoherent background.

\begin{figure}[!t]
    \begin{center}
        \includegraphics[width=\columnwidth]{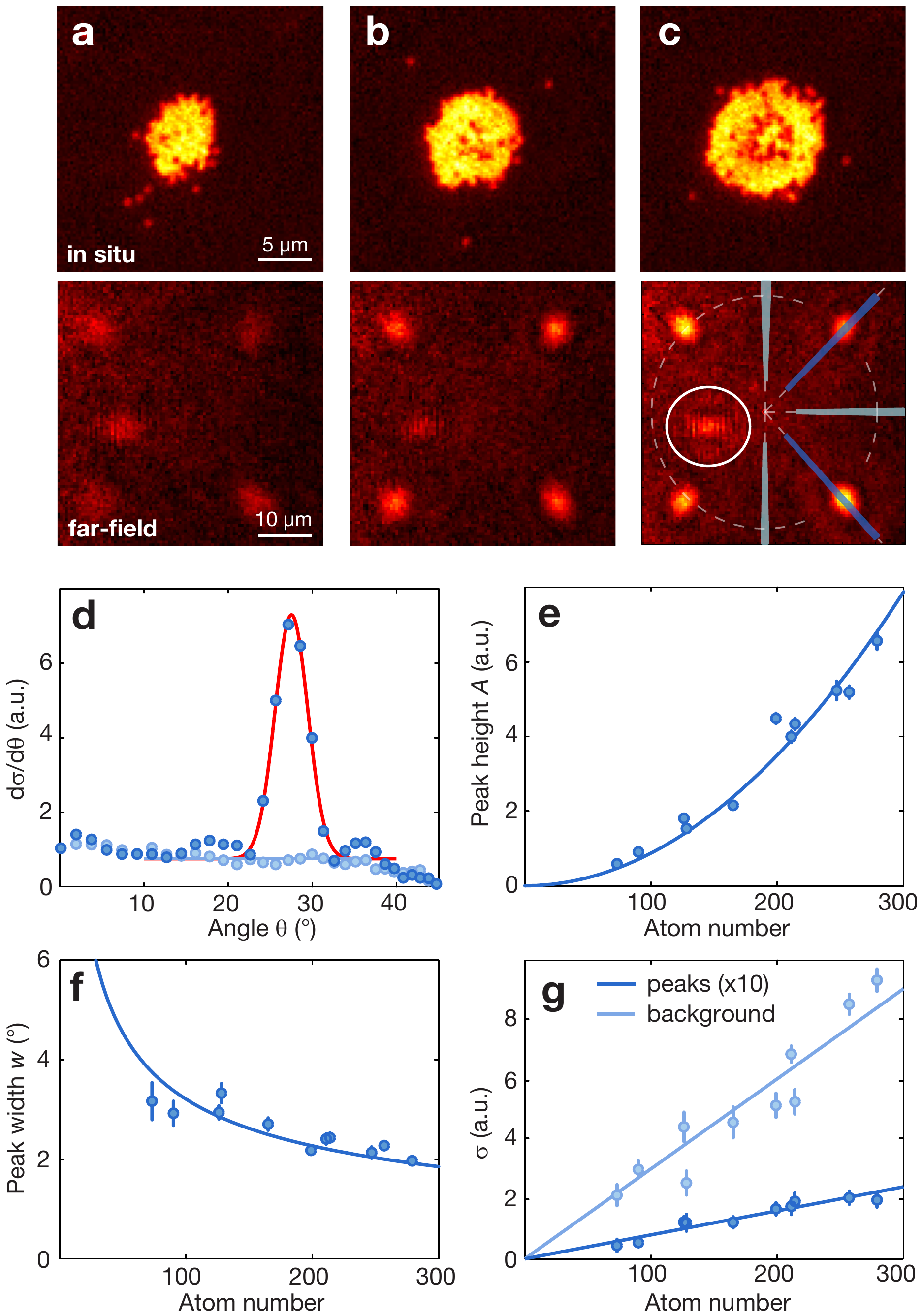}
     \vspace{-0.7cm}
    \end{center}
\caption{Analysis of the diffraction patterns. (a)-(c) In situ images
($N=$126, 199 and 279 atoms) and the corresponding far-field
diffraction pattern ($\Delta z=40\,\mu$m) from the same experimental
run, with illumination time 500\,ms each. (d) Angular distribution
of the differential scattering cross section obtained from cuts in (c)
through the diffraction peaks (dark blue) and the background signal
(light blue). The peak is fitted with a
Gaussian (red line). (e) Peak height $A$ versus $N$ together with a
quadratic fit. (f) Peak width $w$ versus $N$, with a fit to $w\propto 1/
\sqrt{N}$. (g) Total background signal $\sigma_{b}$
(light blue) as obtained from the constant background in (d) and total
signal in the peaks $\sigma_{p}$ (dark blue, scaled by a factor of 10)
obtained from the fits to the diffraction peaks (see
text).\vspace{-0.3cm} \label{fig:PeakEvaluation}}
\end{figure}

For a more quantitative analysis, we recorded diffraction patterns
of MIs for different atom numbers
[Fig.\,\ref{fig:PeakEvaluation}(a)-(c)]. We evaluated cuts (angular
sectors of width $4\,^\circ$) through the diffraction peaks and the
background signal [see Fig.\,\ref{fig:PeakEvaluation}(c)] and applied an appropriate coordinate
transformation in order to obtain the angular distribution
$\frac{\rm{d}\sigma}{\rm{d}\theta}(\theta)$, as shown in
Fig.\,\ref{fig:PeakEvaluation}(d). We fitted the resulting peaks
with a Gaussian (height $A$, $1/\sqrt{\rm{e}}$ width $w$, center position $\theta_{-1}$,
and offset fixed at the background value).
The peak position, averaged
over all experimental runs is $|\theta_{-1}|=27.4(6)\,^\circ$, in excellent
agreement with the expected value of $|\theta_{-1}|=27.8\,^\circ$.
The error is dominated by the systematic uncertainty of $\pm 1\,\mu$m in the
determination of $\Delta z$.
The peak height scales quadratically with the atom number
[Fig.\,\ref{fig:PeakEvaluation}(e)], illustrating the coherent
nature of the diffraction peak. The peak width scales as
$w\propto 1/\sqrt{N}$ [Fig.\,\ref{fig:PeakEvaluation}(f)], in agreement
with the result from the 1D model
[Eq.~(\ref{eq:AngularDependence})], assuming $N_x=\sqrt{N}$ atoms in
one dimension.

There are several mechanisms that lead to a background outside the diffraction peaks.
The first mechanism is the deviation from integer occupation of the lattice sites
caused by density fluctuations in the system \cite{Mekhov:2007}.
The second mechanism is the spread of the atoms
in their potential wells, which reduces the peak height via the Debye-Waller factor.
In our case, the dominant mechanism, however, is inelastic light scattering in the molasses configuration.
We used the far-field images to extract the power scattered into
the detected diffraction peaks and the power scattered into the background.
In the cuts outside the diffraction peaks, we found a constant background
$(\frac{\rm{d}\sigma}{\rm{d}\theta})_{b}$ and calculated the total incoherent scattering cross section
$\sigma_{b}=4\pi(\frac{\rm{d}\sigma}{\rm{d}\theta})_{b}$,
assuming an isotropic intensity distribution.
The total scattering cross section from the five molasses beams
can be estimated as $\sigma_{p}\approx 10 \sin(\theta_{-1}) 2\pi A w^2 \cos(\theta_{-1})$.
The factor $\cos(\theta_{-1})$ accounts for the ellipticity of the
diffraction peaks due to the effective ellipticity of the atomic
cloud, when viewed under the angle $\theta_{-1}$.
Fig.\,\ref{fig:PeakEvaluation}(g) shows $\sigma_{b}$ and $\sigma_{p}$,
which both scale linearly with the number of atoms. From the slopes,
we find a fraction $f_{p}=\sigma_{p}/(\sigma_{b}+\sigma_{p})\approx 3\%$
of the power scattered into the detected peaks.
Only about $20\%$ of the coherently scattered light is diffracted into these peaks,
while the forward scattered part is not detected here.

In addition to
the four diffraction peaks from the horizontal molasses beams, a
fifth weaker peak is clearly visible in the center left part of the
far-field images [see white circle in Fig.\,\ref{fig:PeakEvaluation}(c)].
This peak results from the diffraction of the molasses beam which
is shone in from the direction of the imaging system.
It shows that our single plane of a few hundred
atoms in the optical lattice acts as a ``mirror'' for the incoming
laser beam.

\begin{figure}[t]
    \begin{center}
        \includegraphics[width=0.8\columnwidth]{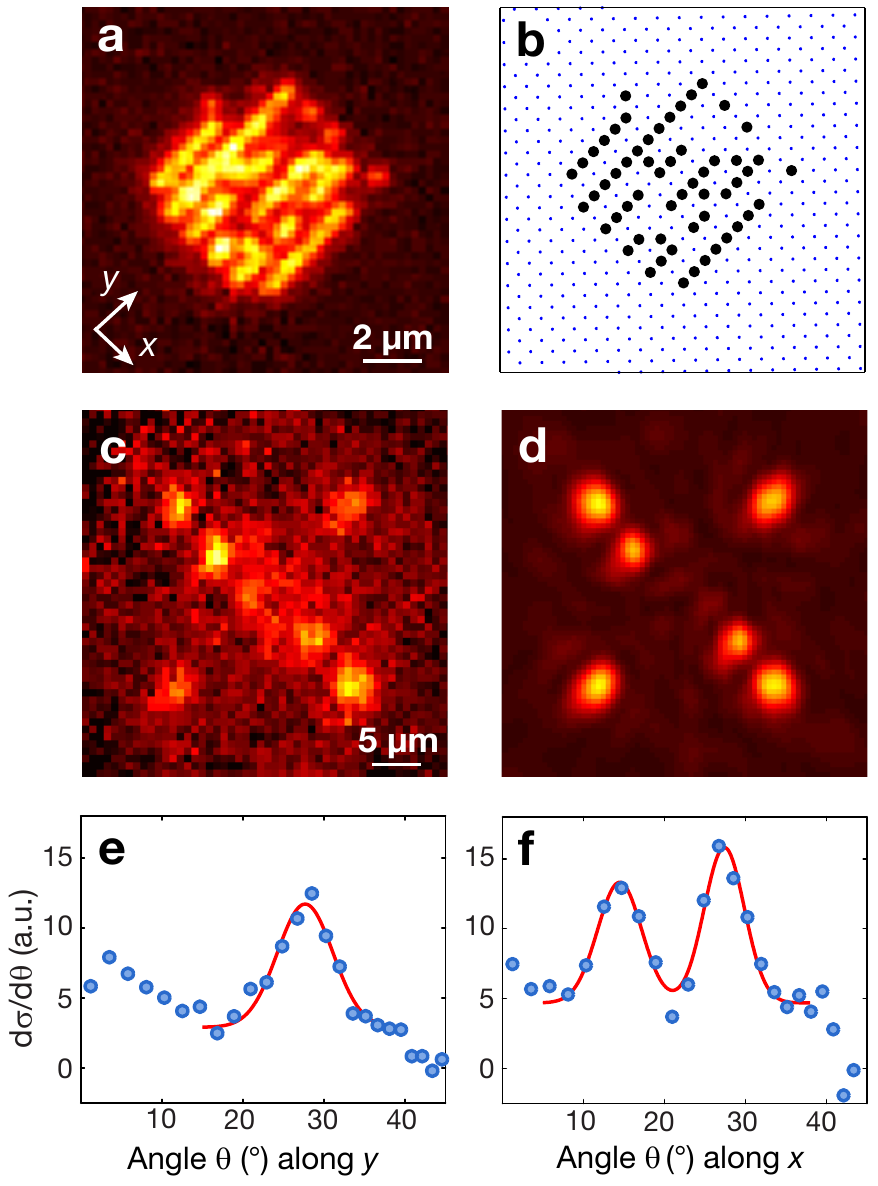}
     \vspace{-0.5cm}
    \end{center}
\caption{Light scattering for a 1D antiferromagnetic order in the density.
(a) The atoms in every second row were removed. (b) Reconstructed atom number distribution from (a).
(c) Resulting far-field image ($\Delta z = 25\um$) with two
diffraction orders in the $x$ direction.  (d) Simulated diffraction
pattern using the atom number distribution of (b). (e),(f) Angular
distribution of the differential scattering cross section obtained
from cuts along the $x$ and $y$ directions together with Gaussian
fits. \vspace{-0.4cm}\label{fig:AFM}}
\end{figure}
Finally, we demonstrated that light scattering can be used for the
detection of spin correlations. As an example, we created a 1D $z$-N\'{e}el
antiferromagnetic order along the $x$ direction of the lattice using our
recently demonstrated single-site addressing technique \cite{Weitenberg:2011}.
We sequentially flipped all atomic spins in every second row of the
lattice from $F=1$ to $F=2$ before we applied a resonant laser that
removed all atoms in $F=2$. Figs.\,\ref{fig:AFM}(a) and (b) show the
resulting fluorescence image in the focal plane together with the
reconstructed atom number distribution. The corresponding
experimental and theoretical diffraction patterns are displayed in
Figs.\,\ref{fig:AFM}(c) and (d). The two predicted diffraction peaks
of -1$^{\rm st}$ and -2$^{\rm nd}$ order along $x$ are clearly
visible in the experimental picture, although our atomic sample
consisted of only 57 atoms. We obtain the usual peak position of
$|\theta_{-1}^y|=27.7(6)^\circ$ along the $y$ direction, whereas the
two peaks along  $x$ are found at $|\theta_{-1}^x|=14.5(6)^\circ$
and $|\theta_{-2}^x|=27.4(6)^\circ$ [see
Figs.\,\ref{fig:AFM}(e),(f)], in good agreement with the expected
values.
We prepared a 1D antiferromagnetic order, because
the additional diffraction peaks that would arise for a 2D antiferromagnetic
order lie outside the opening angle of our imaging system. However,
the position of the diffraction peaks could be varied in a 2D geometry by changing the angle of the incident beams.
An alternative is to use shorter wavelength probe light,
e.g. near-resonant with the $5S-6P$ transition at 420\,nm for $^{87}$Rb.


Our results could be extended to the study of various density
\cite{Mekhov:2007,Lakomy:2009,Rist:2010} or spin
\cite{deVega:2007,Eckert:2008,Corcovilos:2010} correlations in
optical lattices. Most proposals suggest weak non-destructive
probing, which restricts the signal to only a few photons per atom.
In our alternative approach, we projected the correlations onto the
density before the detection by means of an optical molasses, which
yields a signal of thousands of photons per atom.
This is possible for density correlations, as e.g.~the number squeezing
in a MI can be mapped on the mean density by parity projection
\cite{Bakr:2010,Sherson:2010}. For spin correlations, we have
demonstrated the feasibility of removing one spin state and
observing the additional diffraction peaks from the density
structure. This avoids spin selective coupling of the probe light to
the atoms \cite{Corcovilos:2010}, which is incompatible with
simultaneous laser cooling.
In the 3D case, antiferromagnetic order allows scattering from an
additional plane, but it requires careful alignment of the probe beam
angle to match the Bragg condition. In contrast,
light scattering from a 2D system yields additional peaks from the same incident beam,
which is also convenient for the extraction of the spin
correlation length from the relative height or width of the diffraction peaks.
Finally, we note that the detection of spin correlations via light
scattering does not rely on the high aperture of our imaging system.
The diffraction peaks could also be detected in a restricted angular
range.


We acknowledge funding by MPG, DFG, Stiftung Rheinland-Pfalz f\"{u}r
Innovation, Carl-Zeiss Stiftung, EU (NAMEQUAM, AQUTE, Marie Curie
Fellowship to M.C.), and JSPS (Postdoctoral Fellowship for Research
Abroad to T.F.).

\vspace{-0.6cm}
\bibliography{References}

\begin{thebibliography}{20}%
\makeatletter
\providecommand \@ifxundefined [1]{%
 \@ifx{#1\undefined}
}%
\providecommand \@ifnum [1]{%
 \ifnum #1\expandafter \@firstoftwo
 \else \expandafter \@secondoftwo
 \fi
}%
\providecommand \@ifx [1]{%
 \ifx #1\expandafter \@firstoftwo
 \else \expandafter \@secondoftwo
 \fi
}%
\providecommand \natexlab [1]{#1}%
\providecommand \enquote  [1]{``#1''}%
\providecommand \bibnamefont  [1]{#1}%
\providecommand \bibfnamefont [1]{#1}%
\providecommand \citenamefont [1]{#1}%
\providecommand \href@noop [0]{\@secondoftwo}%
\providecommand \href [0]{\begingroup \@sanitize@url \@href}%
\providecommand \@href[1]{\@@startlink{#1}\@@href}%
\providecommand \@@href[1]{\endgroup#1\@@endlink}%
\providecommand \@sanitize@url [0]{\catcode `\\12\catcode `\$12\catcode
  `\&12\catcode `\#12\catcode `\^12\catcode `\_12\catcode `\%12\relax}%
\providecommand \@@startlink[1]{}%
\providecommand \@@endlink[0]{}%
\providecommand \url  [0]{\begingroup\@sanitize@url \@url }%
\providecommand \@url [1]{\endgroup\@href {#1}{\urlprefix }}%
\providecommand \urlprefix  [0]{URL }%
\providecommand \Eprint [0]{\href }%
\providecommand \doibase [0]{http://dx.doi.org/}%
\providecommand \selectlanguage [0]{\@gobble}%
\providecommand \bibinfo  [0]{\@secondoftwo}%
\providecommand \bibfield  [0]{\@secondoftwo}%
\providecommand \translation [1]{[#1]}%
\providecommand \BibitemOpen [0]{}%
\providecommand \bibitemStop [0]{}%
\providecommand \bibitemNoStop [0]{.\EOS\space}%
\providecommand \EOS [0]{\spacefactor3000\relax}%
\providecommand \BibitemShut  [1]{\csname bibitem#1\endcsname}%
\let\auto@bib@innerbib\@empty
\bibitem [{\citenamefont {Bloch}\ \emph {et~al.}(2008)\citenamefont {Bloch},
  \citenamefont {Dalibard},\ and\ \citenamefont {Zwerger}}]{Bloch:2008c}%
  \BibitemOpen
  \bibfield  {author} {\bibinfo {author} {\bibfnamefont {I.}~\bibnamefont
  {Bloch}}, \bibinfo {author} {\bibfnamefont {J.}~\bibnamefont {Dalibard}}, \
  and\ \bibinfo {author} {\bibfnamefont {W.}~\bibnamefont {Zwerger}},\ }\href
  {\doibase 10.1103/RevModPhys.80.885} {\bibfield  {journal} {\bibinfo
  {journal} {Rev. Mod. Phys.}\ }\textbf {\bibinfo {volume} {80}},\ \bibinfo
  {pages} {885} (\bibinfo {year} {2008})}\BibitemShut {NoStop}%
\bibitem [{\citenamefont {Lewenstein}\ \emph {et~al.}(2007)\citenamefont
  {Lewenstein}, \citenamefont {Sanpera}, \citenamefont {Ahufinger},
  \citenamefont {Damski}, \citenamefont {Sen},\ and\ \citenamefont
  {Sen}}]{Lewenstein:2007}%
  \BibitemOpen
  \bibfield  {author} {\bibinfo {author} {\bibfnamefont {M.}~\bibnamefont
  {Lewenstein}}, \bibinfo {author} {\bibfnamefont {A.}~\bibnamefont {Sanpera}},
  \bibinfo {author} {\bibfnamefont {V.}~\bibnamefont {Ahufinger}}, \bibinfo
  {author} {\bibfnamefont {B.}~\bibnamefont {Damski}}, \bibinfo {author}
  {\bibfnamefont {A.}~\bibnamefont {Sen}}, \ and\ \bibinfo {author}
  {\bibfnamefont {U.}~\bibnamefont {Sen}},\ }\href 
  {\doibase 10.1080/00018730701223200} {\bibfield  {journal} {\bibinfo  {journal} {Adv.
  Phys.}\ }\textbf {\bibinfo {volume} {56}},\ \bibinfo {pages} {243} (\bibinfo
  {year} {2007})}\BibitemShut {NoStop}%
\bibitem [{\citenamefont {de~Vega}\ \emph {et~al.}(2008)\citenamefont
  {de~Vega}, \citenamefont {Cirac},\ and\ \citenamefont
  {Porras}}]{deVega:2007}%
  \BibitemOpen
  \bibfield  {author} {\bibinfo {author} {\bibfnamefont {I.}~\bibnamefont
  {de~Vega}}, \bibinfo {author} {\bibfnamefont {J.~I.}\ \bibnamefont {Cirac}},
  \ and\ \bibinfo {author} {\bibfnamefont {D.}~\bibnamefont {Porras}},\ }\href
  {\doibase 10.1103/PhysRevA.77.051804} {\bibfield  {journal} {\bibinfo
  {journal} {Phys. Rev. A}\ }\textbf {\bibinfo {volume} {77}},\ \bibinfo
  {pages} {051804(R)} (\bibinfo {year} {2008})}\BibitemShut {NoStop}%
\bibitem [{\citenamefont {Eckert}\ \emph {et~al.}(2007)\citenamefont {Eckert},
  \citenamefont {Romero-Isart}, \citenamefont {Rodriguez}, \citenamefont
  {Lewenstein}, \citenamefont {Polzik},\ and\ \citenamefont
  {Sanpera}}]{Eckert:2008}%
  \BibitemOpen
  \bibfield  {author} {\bibinfo {author} {\bibfnamefont {K.}~\bibnamefont
  {Eckert}}, \bibinfo {author} {\bibfnamefont {O.}~\bibnamefont
  {Romero-Isart}}, \bibinfo {author} {\bibfnamefont {M.}~\bibnamefont
  {Rodriguez}}, \bibinfo {author} {\bibfnamefont {M.}~\bibnamefont
  {Lewenstein}}, \bibinfo {author} {\bibfnamefont {E.~S.}\ \bibnamefont
  {Polzik}}, \ and\ \bibinfo {author} {\bibfnamefont {A.}~\bibnamefont
  {Sanpera}},\ }\href {\doibase 10.1038/nphys776} {\bibfield  {journal}
  {\bibinfo  {journal} {Nature Physics}\ }\textbf {\bibinfo {volume} {4}},\
  \bibinfo {pages} {50} (\bibinfo {year} {2007})}\BibitemShut {NoStop}%
\bibitem [{\citenamefont {Corcovilos}\ \emph {et~al.}(2010)\citenamefont
  {Corcovilos}, \citenamefont {Baur}, \citenamefont {Hitchcock}, \citenamefont
  {Mueller},\ and\ \citenamefont {Hulet}}]{Corcovilos:2010}%
  \BibitemOpen
  \bibfield  {author} {\bibinfo {author} {\bibfnamefont {T.~A.}\ \bibnamefont
  {Corcovilos}}, \bibinfo {author} {\bibfnamefont {S.~K.}\ \bibnamefont
  {Baur}}, \bibinfo {author} {\bibfnamefont {J.~M.}\ \bibnamefont {Hitchcock}},
  \bibinfo {author} {\bibfnamefont {E.~J.}\ \bibnamefont {Mueller}}, \ and\
  \bibinfo {author} {\bibfnamefont {R.~G.}\ \bibnamefont {Hulet}},\ }\href
  {\doibase 10.1103/PhysRevA.81.013415} {\bibfield  {journal} {\bibinfo
  {journal} {Phys. Rev. A}\ }\textbf {\bibinfo {volume} {81}},\ \bibinfo
  {pages} {013415} (\bibinfo {year} {2010})}\BibitemShut {NoStop}%
\bibitem [{\citenamefont {Ruostekoski}\ \emph {et~al.}(2009)\citenamefont
  {Ruostekoski}, \citenamefont {Foot},\ and\ \citenamefont
  {Deb}}]{Ruostekoski:2009}%
  \BibitemOpen
  \bibfield  {author} {\bibinfo {author} {\bibfnamefont {J.}~\bibnamefont
  {Ruostekoski}}, \bibinfo {author} {\bibfnamefont {C.~J.}\ \bibnamefont
  {Foot}}, \ and\ \bibinfo {author} {\bibfnamefont {A.~B.}\ \bibnamefont
  {Deb}},\ }\href {\doibase 10.1103/PhysRevLett.103.170404} {\bibfield
  {journal} {\bibinfo  {journal} {Phys. Rev. Lett.}\ }\textbf {\bibinfo
  {volume} {103}},\ \bibinfo {pages} {170404} (\bibinfo {year}
  {2009})}\BibitemShut {NoStop}%
\bibitem [{\citenamefont {Mekhov}\ \emph
  {et~al.}(2007{\natexlab{a}})\citenamefont {Mekhov}, \citenamefont
  {Maschler},\ and\ \citenamefont {Ritsch}}]{Mekhov:2007}%
  \BibitemOpen
  \bibfield  {author} {\bibinfo {author} {\bibfnamefont {I.~B.}\ \bibnamefont
  {Mekhov}}, \bibinfo {author} {\bibfnamefont {C.}~\bibnamefont {Maschler}}, \
  and\ \bibinfo {author} {\bibfnamefont {H.}~\bibnamefont {Ritsch}},\ }\href
  {\doibase 10.1103/PhysRevA.76.053618} {\bibfield  {journal} {\bibinfo
  {journal} {Phys. Rev. A}\ }\textbf {\bibinfo {volume} {76}},\ \bibinfo
  {pages} {053618} (\bibinfo {year} {2007}{\natexlab{a}})}\BibitemShut
  {NoStop}%
\bibitem [{\citenamefont {Lakomy}\ \emph {et~al.}(2009)\citenamefont {Lakomy},
  \citenamefont {Idziaszek},\ and\ \citenamefont {Trippenbach}}]{Lakomy:2009}%
  \BibitemOpen
  \bibfield  {author} {\bibinfo {author} {\bibfnamefont {K.}~\bibnamefont
  {Lakomy}}, \bibinfo {author} {\bibfnamefont {Z.}~\bibnamefont {Idziaszek}}, \
  and\ \bibinfo {author} {\bibfnamefont {M.}~\bibnamefont {Trippenbach}},\
  }\href {\doibase 10.1103/PhysRevA.80.043404} {\bibfield  {journal} {\bibinfo
  {journal} {Phys. Rev. A}\ }\textbf {\bibinfo {volume} {80}},\ \bibinfo
  {pages} {043404} (\bibinfo {year} {2009})}\BibitemShut {NoStop}%
\bibitem [{\citenamefont {Rist}\ \emph {et~al.}(2010)\citenamefont {Rist},
  \citenamefont {Menotti},\ and\ \citenamefont {Morigi}}]{Rist:2010}%
  \BibitemOpen
  \bibfield  {author} {\bibinfo {author} {\bibfnamefont {S.}~\bibnamefont
  {Rist}}, \bibinfo {author} {\bibfnamefont {C.}~\bibnamefont {Menotti}}, \
  and\ \bibinfo {author} {\bibfnamefont {G.}~\bibnamefont {Morigi}},\ }\href
  {\doibase 10.1103/PhysRevA.81.013404} {\bibfield  {journal} {\bibinfo
  {journal} {Phys. Rev. A}\ }\textbf {\bibinfo {volume} {81}},\ \bibinfo
  {pages} {013404} (\bibinfo {year} {2010})}\BibitemShut {NoStop}%
\bibitem [{\citenamefont {Mekhov}\ \emph
  {et~al.}(2007{\natexlab{b}})\citenamefont {Mekhov}, \citenamefont
  {Maschler},\ and\ \citenamefont {Ritsch}}]{Mekhov:2007a}%
  \BibitemOpen
  \bibfield  {author} {\bibinfo {author} {\bibfnamefont {I.~B.}\ \bibnamefont
  {Mekhov}}, \bibinfo {author} {\bibfnamefont {C.}~\bibnamefont {Maschler}}, \
  and\ \bibinfo {author} {\bibfnamefont {H.}~\bibnamefont {Ritsch}},\ }\href
  {\doibase 10.1038/nphys571} {\bibfield  {journal} {\bibinfo  {journal}
  {Nature Phys.}\ }\textbf {\bibinfo {volume} {3}},\ \bibinfo {pages} {319}
  (\bibinfo {year} {2007}{\natexlab{b}})}\BibitemShut {NoStop}%
\bibitem [{\citenamefont {Chen}\ \emph {et~al.}(2007)\citenamefont {Chen},
  \citenamefont {Meiser},\ and\ \citenamefont {Meystre}}]{Chen:2007}%
  \BibitemOpen
  \bibfield  {author} {\bibinfo {author} {\bibfnamefont {W.}~\bibnamefont
  {Chen}}, \bibinfo {author} {\bibfnamefont {D.}~\bibnamefont {Meiser}}, \ and\
  \bibinfo {author} {\bibfnamefont {P.}~\bibnamefont {Meystre}},\ }\href
  {\doibase 10.1103/PhysRevA.75.023812} {\bibfield  {journal} {\bibinfo
  {journal} {Phys. Rev. A}\ }\textbf {\bibinfo {volume} {75}},\ \bibinfo
  {pages} {023812} (\bibinfo {year} {2007})}\BibitemShut {NoStop}%
\bibitem [{\citenamefont {Birkl}\ \emph {et~al.}(1995)\citenamefont {Birkl},
  \citenamefont {Gatzke}, \citenamefont {Deutsch}, \citenamefont {Rolston},\
  and\ \citenamefont {Phillips}}]{Birkl:1995}%
  \BibitemOpen
  \bibfield  {author} {\bibinfo {author} {\bibfnamefont {G.}~\bibnamefont
  {Birkl}}, \bibinfo {author} {\bibfnamefont {M.}~\bibnamefont {Gatzke}},
  \bibinfo {author} {\bibfnamefont {I.~H.}\ \bibnamefont {Deutsch}}, \bibinfo
  {author} {\bibfnamefont {S.~L.}\ \bibnamefont {Rolston}}, \ and\ \bibinfo
  {author} {\bibfnamefont {W.~D.}\ \bibnamefont {Phillips}},\ }\href 
  {\doibase 10.1103/PhysRevLett.75.2823} {\bibfield  {journal} {\bibinfo  {journal}
  {Phys. Rev. Lett.}\ }\textbf {\bibinfo {volume} {75}},\ \bibinfo {pages}
  {2823} (\bibinfo {year} {1995})}\BibitemShut {NoStop}%
\bibitem [{\citenamefont {Weidem\"{u}ller}\ \emph {et~al.}(1995)\citenamefont
  {Weidem\"{u}ller}, \citenamefont {Hemmerich}, \citenamefont {G\"{o}rlitz},
  \citenamefont {Esslinger},\ and\ \citenamefont
  {H\"{a}nsch}}]{Weidemueller:1995}%
  \BibitemOpen
  \bibfield  {author} {\bibinfo {author} {\bibfnamefont {M.}~\bibnamefont
  {Weidem\"{u}ller}}, \bibinfo {author} {\bibfnamefont {A.}~\bibnamefont
  {Hemmerich}}, \bibinfo {author} {\bibfnamefont {A.}~\bibnamefont
  {G\"{o}rlitz}}, \bibinfo {author} {\bibfnamefont {T.}~\bibnamefont
  {Esslinger}}, \ and\ \bibinfo {author} {\bibfnamefont {T.~W.}~\bibnamefont
  {H\"{a}nsch}},\ }\href {\doibase 10.1103/PhysRevLett.75.4583} {\bibfield
  {journal} {\bibinfo  {journal} {Phys. Rev. Lett.}\ }\textbf {\bibinfo
  {volume} {75}},\ \bibinfo {pages} {4583} (\bibinfo {year}
  {1995})}\BibitemShut {NoStop}%
\bibitem [{\citenamefont {Weidem\"{u}ller}\ \emph {et~al.}(1998)\citenamefont
  {Weidem\"{u}ller}, \citenamefont {G\"{o}rlitz}, \citenamefont {H\"{a}nsch},\
  and\ \citenamefont {Hemmerich}}]{Weidemueller:1998}%
  \BibitemOpen
  \bibfield  {author} {\bibinfo {author} {\bibfnamefont {M.}~\bibnamefont
  {Weidem\"{u}ller}}, \bibinfo {author} {\bibfnamefont {A.}~\bibnamefont
  {G\"{o}rlitz}}, \bibinfo {author} {\bibfnamefont {T.~W.}\ \bibnamefont
  {H\"{a}nsch}}, \ and\ \bibinfo {author} {\bibfnamefont {A.}~\bibnamefont
  {Hemmerich}},\ }\href {\doibase 10.1103/PhysRevA.58.4647} {\bibfield
  {journal} {\bibinfo  {journal} {Phys. Rev. A}\ }\textbf {\bibinfo {volume}
  {58}},\ \bibinfo {pages} {4647} (\bibinfo {year} {1998})}\BibitemShut
  {NoStop}%
\bibitem [{\citenamefont {Westbrook}\ \emph {et~al.}(1997)\citenamefont
  {Westbrook}, \citenamefont {Jurczak}, \citenamefont {Birkl}, \citenamefont
  {Desruelle}, \citenamefont {Phillips},\ and\ \citenamefont
  {Aspect}}]{Westbrook:1997}%
  \BibitemOpen
  \bibfield  {author} {\bibinfo {author} {\bibfnamefont {C.~I.}\ \bibnamefont
  {Westbrook}}, \bibinfo {author} {\bibfnamefont {C.}~\bibnamefont {Jurczak}},
  \bibinfo {author} {\bibfnamefont {G.}~\bibnamefont {Birkl}}, \bibinfo
  {author} {\bibfnamefont {B.}~\bibnamefont {Desruelle}}, \bibinfo {author}
  {\bibfnamefont {W.~D.}\ \bibnamefont {Phillips}}, \ and\ \bibinfo {author}
  {\bibfnamefont {A.}~\bibnamefont {Aspect}},\ }\href 
  {\doibase 10.1080/09500349708231850} {\bibfield  {journal} {\bibinfo  {journal} {J.
  Mod. Opt.}\ }\textbf {\bibinfo {volume} {44}},\ \bibinfo {pages} {1837}
  (\bibinfo {year} {1997})}\BibitemShut {NoStop}%
\bibitem [{\citenamefont {Raithel}\ \emph {et~al.}(1997)\citenamefont
  {Raithel}, \citenamefont {Birkl}, \citenamefont {Kastberg}, \citenamefont
  {Phillips},\ and\ \citenamefont {Rolston}}]{Raithel:1997}%
  \BibitemOpen
  \bibfield  {author} {\bibinfo {author} {\bibfnamefont {G.}~\bibnamefont
  {Raithel}}, \bibinfo {author} {\bibfnamefont {G.}~\bibnamefont {Birkl}},
  \bibinfo {author} {\bibfnamefont {A.}~\bibnamefont {Kastberg}}, \bibinfo
  {author} {\bibfnamefont {W.~D.}\ \bibnamefont {Phillips}}, \ and\ \bibinfo
  {author} {\bibfnamefont {S.~L.}\ \bibnamefont {Rolston}},\ }\href 
  {\doibase 10.1103/PhysRevLett.78.630} {\bibfield  {journal} {\bibinfo  {journal} {Phys.
  Rev. Lett.}\ }\textbf {\bibinfo {volume} {78}},\ \bibinfo {pages} {630}
  (\bibinfo {year} {1997})}\BibitemShut {NoStop}%
\bibitem [{\citenamefont {Slama}\ \emph {et~al.}(2005)\citenamefont {Slama},
  \citenamefont {von Cube}, \citenamefont {Deh}, \citenamefont {Ludewig},
  \citenamefont {Zimmermann},\ and\ \citenamefont {Courteille}}]{Slama:2005a}%
  \BibitemOpen
  \bibfield  {author} {\bibinfo {author} {\bibfnamefont {S.}~\bibnamefont
  {Slama}}, \bibinfo {author} {\bibfnamefont {C.}~\bibnamefont {von Cube}},
  \bibinfo {author} {\bibfnamefont {B.}~\bibnamefont {Deh}}, \bibinfo {author}
  {\bibfnamefont {A.}~\bibnamefont {Ludewig}}, \bibinfo {author} {\bibfnamefont
  {C.}~\bibnamefont {Zimmermann}}, \ and\ \bibinfo {author} {\bibfnamefont
  {P.~W.}~\bibnamefont {Courteille}},\ }\href 
  {\doibase 10.1103/PhysRevLett.94.193901} {\bibfield  {journal} {\bibinfo  {journal}
  {Phys. Rev. Lett.}\ }\textbf {\bibinfo {volume} {94}},\ \bibinfo {pages}
  {193901} (\bibinfo {year} {2005})}\BibitemShut {NoStop}%
\bibitem [{\citenamefont {Sherson}\ \emph {et~al.}(2010)\citenamefont
  {Sherson}, \citenamefont {Weitenberg}, \citenamefont {Endres}, \citenamefont
  {Cheneau}, \citenamefont {Bloch},\ and\ \citenamefont {Kuhr}}]{Sherson:2010}%
  \BibitemOpen
  \bibfield  {author} {\bibinfo {author} {\bibfnamefont {J.~F.}\ \bibnamefont
  {Sherson}}, \bibinfo {author} {\bibfnamefont {C.}~\bibnamefont {Weitenberg}},
  \bibinfo {author} {\bibfnamefont {M.}~\bibnamefont {Endres}}, \bibinfo
  {author} {\bibfnamefont {M.}~\bibnamefont {Cheneau}}, \bibinfo {author}
  {\bibfnamefont {I.}~\bibnamefont {Bloch}}, \ and\ \bibinfo {author}
  {\bibfnamefont {S.}~\bibnamefont {Kuhr}},\ }\href 
  {\doibase 10.1038/nature09378} {\bibfield  {journal} {\bibinfo  {journal} {Nature}\
  }\textbf {\bibinfo {volume} {467}},\ \bibinfo {pages} {68} (\bibinfo {year}
  {2010})}\BibitemShut {NoStop}%
\bibitem [{\citenamefont {Weitenberg}\ \emph {et~al.}(2011)\citenamefont
  {Weitenberg}, \citenamefont {Endres}, \citenamefont {Sherson}, \citenamefont
  {Cheneau}, \citenamefont {Schau\ss}, \citenamefont {Fukuhara}, \citenamefont
  {Bloch},\ and\ \citenamefont {Kuhr}}]{Weitenberg:2011}%
  \BibitemOpen
  \bibfield  {author} {\bibinfo {author} {\bibfnamefont {C.}~\bibnamefont
  {Weitenberg}}, \bibinfo {author} {\bibfnamefont {M.}~\bibnamefont {Endres}},
  \bibinfo {author} {\bibfnamefont {J.~F.}\ \bibnamefont {Sherson}}, \bibinfo
  {author} {\bibfnamefont {M.}~\bibnamefont {Cheneau}}, \bibinfo {author}
  {\bibfnamefont {P.}~\bibnamefont {Schau\ss}}, \bibinfo {author}
  {\bibfnamefont {T.}~\bibnamefont {Fukuhara}}, \bibinfo {author}
  {\bibfnamefont {I.}~\bibnamefont {Bloch}}, \ and\ \bibinfo {author}
  {\bibfnamefont {S.}~\bibnamefont {Kuhr}},\ }\href
  {http://arxiv.org/abs/1101.2076} {\bibfield  {journal} {\bibinfo  {journal}
  {arXiv:1101.2076v1}\ } (\bibinfo {year} {2011})}\BibitemShut {NoStop}%
\bibitem [{\citenamefont {Bakr}\ \emph {et~al.}(2010)\citenamefont {Bakr},
  \citenamefont {Peng}, \citenamefont {Tai}, \citenamefont {Ma}, \citenamefont
  {Simon}, \citenamefont {Gillen}, \citenamefont {F\"{o}lling}, \citenamefont
  {Pollet},\ and\ \citenamefont {Greiner}}]{Bakr:2010}%
  \BibitemOpen
  \bibfield  {author} {\bibinfo {author} {\bibfnamefont {W.~S.}\ \bibnamefont
  {Bakr}}, \bibinfo {author} {\bibfnamefont {A.}~\bibnamefont {Peng}}, \bibinfo
  {author} {\bibfnamefont {M.~E.}\ \bibnamefont {Tai}}, \bibinfo {author}
  {\bibfnamefont {R.}~\bibnamefont {Ma}}, \bibinfo {author} {\bibfnamefont
  {J.}~\bibnamefont {Simon}}, \bibinfo {author} {\bibfnamefont {J.~I.}\
  \bibnamefont {Gillen}}, \bibinfo {author} {\bibfnamefont {S.}~\bibnamefont
  {F\"{o}lling}}, \bibinfo {author} {\bibfnamefont {L.}~\bibnamefont {Pollet}},
  \ and\ \bibinfo {author} {\bibfnamefont {M.}~\bibnamefont {Greiner}},\ }\href
  {\doibase 10.1126/science.1192368} {\bibfield  {journal} {\bibinfo  {journal}
  {Science}\ }\textbf {\bibinfo {volume} {329}},\ \bibinfo {pages} {547}
  (\bibinfo {year} {2010})}\BibitemShut {NoStop}%
\end{thebibliography}%


\end{document}